\documentstyle[12pt,psfig]{article}

\everyjob{\typeout{MuTeX.sty: LaTeX Modification by EPM.}}
\immediate\write10{MuTeX.sty: LaTeX Modification by EPM.}

\makeatletter   

\input{ifthen.sty}

\newtheorem {theorem} {Theorem} [section]

\newtheorem {Canham threshold} [theorem] {Canham Threshold}

\def\theoremstyle#1#2{\def\@@theoremheadstyle{#1}
                      \def\@@theorembodystyle{#2}}
\def\@@theoremheadstyle{\sc}
\def\@@theorembodystyle{\rm}
\def\@begintheorem#1#2{\@@theorembodystyle 
                       \trivlist 
		       \item[\hskip 
                             \labelsep{\@@theoremheadstyle #1\ #2}]}
\def\@opargbegintheorem#1#2#3{\@@theorembodystyle 
                              \trivlist 
			       \item[\hskip 
				  \labelsep{\@@theoremheadstyle #1\ #2\ (#3)}]}

 \def\@@pc{\bf}

 \newcommand {\pcodestyle}[1] {\def\@@pc{#1}}  

 {
 \def\PROGRAM		{{\@@pc program\ }}
 
 \def\PROCEDURE		{{\@@pc procedure\ }}
 \def\FUNCTION		{{\@@pc function\ }}
 \def\LOCAL		{{\@@pc local\ }}
 \def\GLOBAL		{{\@@pc global\ }}
 \def\RETURNS		{{\@@pc returns\ }}
 \def\RETURN		{{\@@pc return\ }}
 \def\BEGIN		{{\@@pc begin\ }}
 \def\END		{{\@@pc end\ }}
 \def\IF			{{\@@pc if\ }}
 \def\THEN		{{\@@pc then\ }}
 \def\ELSE		{{\@@pc else\ }}
 \def\REPEAT		{{\@@pc repeat\ }}
 \def\UNTIL		{{\@@pc until\ }}
 \def\WHILE		{{\@@pc while\ }}
 \def\DO			{{\@@pc do\ }}
 \def\FOR		{{\@@pc for\ }}
 \def\TO			{{\@@pc to\ }}
 \def\DOWN		{{\@@pc down\ }}
 \def\NEXT		{{\@@pc next\ }}
 \begin{center}
  \begin{minipage}{.9\textwidth}
   \small
    \begin{tabbing}
 }%
 {  \end{tabbing}%
   \end{minipage}%
  \end{center}%
 }

			   {\end{quotation}}

			 {  \end{minipage}%
			   \end{center}%
			  \end{description}%
			 }

				{ \end{minipage}
				 \end{center}%
				}

\def\thebibliography#1{\section*{References}\list
 {[\arabic{enumi}]}{\settowidth\labelwidth{[#1]}\leftmargin\labelwidth
 \advance\leftmargin\labelsep
 \usecounter{enumi}}
 \def\newblock{\hskip .11em plus .33em minus -.07em}
 \sloppy
 \sfcode`\.=1000\relax}

\newsavebox{\ProofSym}
\savebox{\ProofSym}{%
  \begin{picture}(10,10)
    \put(0,0){\framebox(9,9){}}
    \put(0,3){\framebox(6,6){}}
  \end{picture}}
\newcommand{\eop}{\hfill\usebox{\ProofSym}}

\begin{document}
\title{Characterizing the combinatorics of distributed EPR pairs for 
multi-partite entanglement}
\author{Sudhir Kumar Singh
\and Somesh Kumar\\emails:(sudhirks, smsh)@maths.iitkgp.ernet.in\\
Department of Mathematics\\
Indian Institute of Technology, Kharagpur, 721302, India\\
\and Sudebkumar Prasant Pal\\email:spp@cse.iitkgp.ernet.in\\
http://www.facweb.iitkgp.ernet.in/\~~spp/\\
Department of Computer Science and Engineering\\
Indian Institute of Technology, Kharagpur, 721302,
India \\}

\date{}
\maketitle

\begin{abstract}
We develop protocols for preparing a GHZ state and, in general, 
a pure multi-partite maximally entangled state in a distributed network 
with apriori quantum entanglement between agents 
using classical communication and  
local operations. We investigate and characterize the minimal 
combinatorics of the sharing of EPR pairs required amongst agents 
in a network for the creation of multi-partite entanglement. 
We also characterize the minimal combinatorics 
of agents in the creation of pure maximal multi-partite 
entanglement amongst the set $N$ of $n$ agents in a network 
using apriori multi-partite entanglement states amongst subsets of 
$N$. We propose protocols for establishing multi-partite entanglement
in the above cases. 
\end{abstract}


\section{Introduction}
\label{intro}

\noindent Quantum entanglement is one of the most remarkable aspects 
of quantum physics. Two particles in an entangled state
behave in some respects as a single entity 
even if the particles are physically separated by a great distance.
The entangled particles
exhibit what physicists call non-local effects. Such non-local effects
were alluded to in the famous 1935 paper by Einstein, Podolsky, and Rosen
\cite{epr1935} and were later referred to as spooky actions at a distance
by Einstein \cite{einstein71}. In 1964, Bell \cite{bell64} formalized the notion of 
two-particle non-locality in terms of correlations amongst probabilities in a
scenario where measurements are performed on each 
particle. He showed that the results of the measurements that occur 
quantum physically can be correlated in a way that cannot occur 
classically unless the type of measurement selected to be performed on
one particle affects the result of the measurement performed on the 
other particle. The two particles thus correlated maximally are called
EPR pairs or Bell pairs. 
Non-local effects, however, without being supplemented by additional quantum or
classical communication, do not convey any signal and therefore
the question of faster than light 
communication does not arise.\\  

Entanglement is a key resource for quantum information processing and
spatially separated entangled pairs of particles have been
used for numerous purposes like teleportation \cite{bennett93}, 
superdense coding \cite{Bwiesner92} and cryptography based on Bell's 
Theorem \cite{ekert91}, to name a few. An EPR channel (a bipartite 
maximally entangled distributed pair of entangled particles) can 
be used in conjunction with a classical communication channel to
send an unknown quantum state of a particle to a distant particle.
The original unknown quantum state is destroyed in the process 
and reproduced at the other end. The process does not copy the original
state; it transports a state and thus does not violate the quantum 
no-cloning theorem \cite{wooters82}. 
This process is called teleportation and was proposed by 
Bennett et al. in their seminal work \cite{bennett93}. 
In teleporation, a quantum communication channel is 
simulated using a classical channel and an EPR channel. A classical 
channel can also be simulated using a quantum channel and an EPR channel
using {\it superdense coding} as 
proposed by Bennett and Wiesner \cite{Bwiesner92}.
Two cbits (classical bits) are compressed in to a qubit and 
sent through an EPR channel to the distant party, who then
recovers the cbits by local operations.  \\

The use of EPR pairs for cryptography was first proposed by Ekert in 1991 
\cite{ekert91}. He proposed a protocol based  
on generalized Bell's theorem for quantum key distribution between 
two parties. The two parties share an EPR pair in advance. They do a
computational basis measurement on their respective qubits and the 
mesurement result is then used as the one bit shared key. 
While the measurement result is maximally uncertain, the 
correlation between their results is deterministic. 
Based on similar principles, a multiparty quantum key 
distribution protocol using EPR pairs in a distributed network 
and its proof of unconditional security has been proposed by 
Singh and Srikanth \cite{sinsrik031}. 
Apart from these applications, entanglement 
has been used in several other
applications such as cheating bit commitment \cite{lochau97}, 
broadcasting of entanglement \cite{buzek97} and testing Bell's 
inequalities \cite{bell64,clauser69,gisin92}.\\
 
Just as two distant particles could be entangled forming an EPR pair,
it is also possible to entangle three or more separated particles. One 
example (called GHZ state) is due to Greenberger, Horne and Zeilinger 
\cite{ghz89}; here
three particles are entangled. A well-known manifestation of 
multipartite entanglement is in testing nonlocality from different 
directions \cite{ghz89, mermin90, home95, peres97}.
Recently, it has also been used for many multi-party computation and 
communication tasks \cite{buhrman01,buhrman99,brassard03,grover97,pal03}
and multi-party cryptography 
\cite{hillary99, scarani01, bose98, choi03, sinsrik032} and \cite{sinsrik033}.
Buhrman, Cleve
and van Dam \cite{buhrman01}, make use of  three-party entanglement 
and demostrate the existence of a function whose computation requires 
strictly lesser classical communication complexity compared to the 
scenario where no entanglement is used. Brassard et al. in 
\cite{brassard03}, show that prior multipartite 
entanglement can be used by
$n$ agents to solve a multi-party distributed problem, whereas no 
classical deterministic protocol succeeds in solving the same 
problem with probability away from half by a fraction that is larger
than an inverse exponential in the number of agents. Buhrman et al.
\cite{buhrman99} solves an $n$-party problem which shows a separation 
of $n$ versus  $\Theta (n\log n)$ bits between quantum and 
classical communication complexity. For one round, three-party problem
this article also proved a difference of $(n+1)$ versus $((3/2)n+1)$ bits
between communication with and without intial entanglement. 
In \cite{pal03}, Pal et al. present a {\it fair} and {\it unbiased} 
leader election protocol using maximal multi-partite entanglement. \\

Quantum teleportation strikingly underlines the peculiar features of
the quantum world. All the properties of the quantum state being 
transfered are retained during this process. So it is natural to ask 
whether one qubit of an entangled state can be teleported while 
retaining its entanglement with the other qubit. The answer is, not
surprisingly, in the affirmative. This is called {\it entanglement swapping}.
Yurke and Stoler \cite{yurke92} and Zukowski et al \cite{zukowski93}
have shown that through entanglement 
swapping one can entangle particles that do not even share any 
common past. This idea has been generalized to the tripartite case by
Zukowski et al. in \cite{zukowski95} and later to the multipartite
case by Zeilinger et al. \cite{zeilinger97} and 
Bose et al. \cite{bose98}. 
Zeilinger et al. presented a general scheme and realizable procedures
for generating GHZ states out of two pairs of entangled particles 
from independent emissions. They also proposed a scheme for observing 
four-particle GHZ state and their scheme can directly be generalized
to more particles. To create a maximally entangled state of $(n+m-1)$ 
particles from two groups, one of $n$ maximally entangled particles 
and the other of $m$ 
maximally entangled particles, it is enough to perform a controlled 
operation between a particle from the first group and 
a particle from the 
second group and then a measurement of the target particle. We observe that 
if particles are distributed in a network then a single 
cbit of communication is required
for broadcasting the mesurement result to construct the 
desired $(n+m-1)$ maximally entangled state using local operations.
In \cite{bose98},
Bose et al. have generalized the entangled swapping scheme of 
Zukowski et al. in a different way. In their scheme, the basic 
ingredient is a projection onto a maximally entangled state of $N$
particles. Each of the $N$ users needs
to share a Bell pair with a central exchange. The central exchange 
then projects the $N$ qubits with it on to an $N$ particle maximally 
enatngled basis, thus leaving the $N$ users in an $N$ partite maximally 
entangled state. However, we note that in order 
to get a desired state, the measurement
result obtained by the central exchange must be broadcast so that
the end users can appropriatly apply requisite local operations. This 
involves $N$ cbits of communication.\\   

In this paper, we consider the problem of creating pure maximally
entangled multi-partite states out of Bell pairs distributed in a
communication network from a physical as well as a 
combinatorial perspective. We 
investigate and characterize
the minimal combinatorics of the distribution
of Bell pairs and show how this combinatorics gives rise to 
resource minimization in long-distance quantum communication.
We present protocols for creating maximal multi-partite entanglement. 
The first protocol (see Theorem \ref{ghztheorem}) enables us to prepare
a GHZ state using two Bell pairs shared amongst the three agents
with the help of two cbits of communication and local operations with 
the additional feature that this 
protocol involves all the three agents {\it dynamically}. Such a protocol with
local dynamic involvement in creating entanglement  
may find applications in cryptographic tasks.
The second protocol (see Theorem \ref{qhtheorem3}) 
entails the use of $O(n)$ cbits of communication and local
operations to prepare a pure $n$-partite maximally entangled state in a
distributed network of Bell pairs; the requirement here is
that the pairs of nodes sharing EPR pairs must form a connected graph. 
We show that a spanning 
tree structure (see Theorem \ref{qhtheorem2}) 
is the minimal combinatorial requirement for creating
multi-partite entanglement. 
We also characterize the minimal combinatorics 
of agents in the creation of pure maximal multi-partite 
entanglement amongst the set $N$ of $n$ agents in a network 
using apriori multi-partite entanglement states amongst subsets of 
$N$. This is done by generalizing the {\it  EPR graph} representation 
to an {\it entangled hypergraph} and the requirement here is that the 
entangled hypergraph representing the entanglement structure must be
connected.\\

This paper is organized as follows. In 
Section \ref{ghzsection}, we present
our protocol I to prepare a GHZ state from two Bell pairs involving all
the three agents dynamically and compare our protocol 
in the light of existing schemes. Section \ref{eprsection} is devoted to
characterizing the spanning tree combinatorics 
of Bell pairs for preparing a pure $n$-partite
maximally entangled state. We develop our protocol II for this purpose. 
In Section \ref{hypersection} we generalize the results of 
Section \ref{eprsection} to the setting where subsets 
of agents in the network
share apriori pure multi-partite maximally entangled states.
Finally in Section \ref{conclusion} we compare our scheme of 
Section \ref{eprsection} with
the multipartite entanglement swapping scheme of Bose et al. \cite{bose98},
observe the similarity
between Helly-type theorems and the combinatorics developed
in Sections \ref{eprsection} and \ref{hypersection} and conclude with 
a few remarks on open research directions.

\section{Preparing a GHZ state from two EPR pairs shared
amongst three agents}
\label{ghzsection}

In this section we consider the preparation of a 
GHZ state from two EPR pairs shared amongst three agents in a 
distributed network. We establish the following theorem. 

\begin{figure}[htbp]
\centerline{\psfig{figure=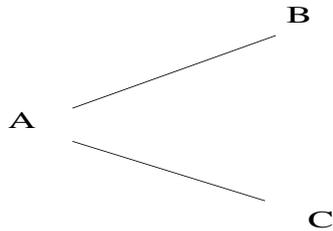,height=4cm,width=5cm}}
\caption{$A$ shares an EPR pair with each of $B$ and $C$.}
\label{figure1}
\end{figure}

\begin{theorem}
If any two pairs of the three agents $A$ (Alice), $B$ (Bob) and $C$
(Charlie) share EPR pairs (say the state $(|00\rangle +
|11\rangle)/\surd{2}$) then we can prepare a GHZ state
$(|000\rangle + |111\rangle)/\surd{2}$ amongst them with 
two bits of classical communication, while involving all the 
three agents dynamically.
\label{ghztheorem}
\end{theorem}

\noindent {Proof:}
The proof follows from the Protocol I.

\noindent {\it The Protocol I:} Without loss of generality let
us assume that the sharing arrangement is as in Figure \ref{figure1}.
$A$ shares an EPR pair with $B$ and another EPR pair with $C$
but $B$ and $C$ do not share an EPR pair. This means that we have the
states $(|0_{a1}0_b\rangle + |1_{a1}1_b\rangle)/\surd{2}$ and
$(|0_{a2}0_c\rangle + |1_{a2}1_c\rangle)/\surd{2}$ where
subscripts $a1$ and $a2$ denote the first and second qubits with $A$ and
subscripts $b$ and $c$ denote qubits with $B$ and $C$, respectively.

\begin{figure}[htbp]
\centerline{\psfig{figure=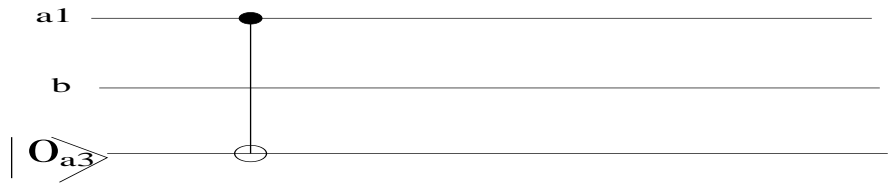,height=8cm,width=12cm}}
\caption{Entangling qubits $a3$ with the EPR pair between $A$ and $B$.}
\label{figure2}
\end{figure}

Our aim is to prepare $(|0_{a1}0_b0_c\rangle +
|1_{a1}1_b1_c\rangle)/\surd{2}$ or $(|0_{a2}0_b0_c\rangle +
|1_{a2}1_b1_c\rangle)/\surd{2}$. We need three steps to do so.

\noindent {\it Step 1:} $A$ prepares a third qubit in the state
$|0\rangle$. We denote this state as $|0_{a3}\rangle$ where the
subscript $a3$ indicates that this is the third qubit of $A$.

\noindent {\it Step 2:} $A$ prepares the state
$(|0_{a1}0_b0_{a3}\rangle +
|1_{a1}1_b1_{a3}\rangle)/\surd{2}$ using 
the circuit in Figure \ref{figure2}.

\noindent {\it Step 3:} $A$ sends her third qubit to $C$ with the help
of the EPR channel $(|0_{a2}0_c\rangle + |1_{a2}1_c\rangle)/\surd{2}$.
A straight forward way to do this is through teleportation. This 
method however, does not involve both of $B$ 
and $C$ dynamically. By a party being dynamic we mean that the party is 
involved in applying the local operations for the completion of 
the transfer of the state of third qubit to create 
the desired GHZ state.

\begin{figure}[htbp]
\centerline{\psfig{figure=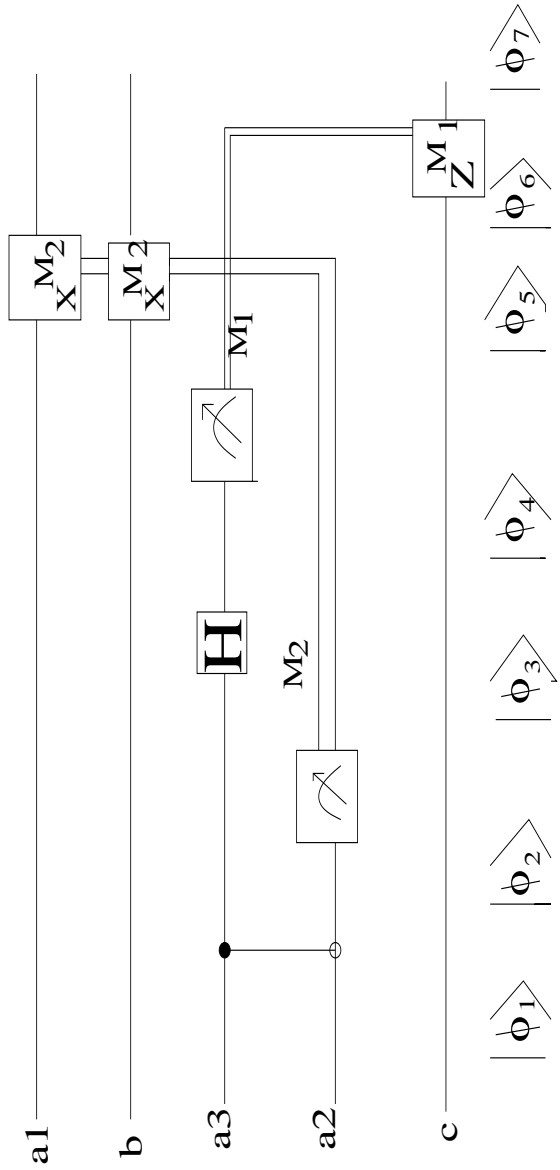,height=18cm,width=10cm}}
\caption{Circuit for creating a GHZ state from two EPR pairs 
with dynamic involment of both $B$ and $C$.}
\label{figure3}
\end{figure}

We use our new and novel teleportation circuit 
as shown in Figure \ref{figure3} where both $B$ and $C$ are dynamic.
The circuit works as follows.
$A$ has all her three qubits with her and can do any
operation she wants to be performed on them. Initially the five
qubits are jointly in the state $|\phi_1\rangle$. $A$ first applies
a controlled NOT gate on her second qubit controlling it from her
third qubit changing $|\phi_1\rangle$ to $|\phi_2\rangle$. Then
she measures her second qubit yielding measurement result $M_2$
and bringing the joint state to $|\phi_3\rangle$. She then applies
a Hadamard gate on her third qubit and the joint state becomes
$|\phi_4\rangle$. A measurement on the third qubit is then done by
her yielding the result $M_1$ and bringing the joint state to
$|\phi_5\rangle$. She then applies a NOT (Pauli's X operator) on
her first qubit, if $M_2$ is 1.  Now she sends the measurement
results $M_2$ to $B$ and $M_1$ to $C$. $B$ applies an $X$ gate on his
qubit if he gets 1 and $C$ applies a $Z$ gate (Pauli's $Z$ operator) if
he gets 1. The order in which $B$ and $C$ apply their operations does
not matter. The final state is $|\phi_7\rangle$. 
The circuit indeed produces the GHZ state between $A$, $B$ and
$C$ as can be seen from the detailed mathematical explanation 
of the circuit given below. It can be noted that this protocol 
requires two cbits of communication.

The above circuit can be explained as
follows:

$$|\phi_1\rangle  = (|0_{a1}0_b0_{a3}\rangle +
|1_{a1}1_b1_{a3}\rangle) (|0_{a2}0_c\rangle +
|1_{a2}1_c\rangle)/2,$$

$$|\phi_2\rangle  = [|0_{a1}0_b0_{a3}\rangle
(|0_{a2}0_c\rangle +
|1_{a2}1_c\rangle) + |1_{a1}1_b1_{a3}\rangle
(|1_{a2}0_c\rangle +
|0_{a2}1_c\rangle)]/2.$$

\noindent {\it Case 1:} $M_2 = 0$
\begin{eqnarray*}
|\phi_3\rangle & = & (|0_{a1}0_b0_{a3}0_{a2}0_c\rangle +
|1_{a1}1_{b}1_{a3}0_{a2}1_c\rangle)/\surd{2}\\
& = & (|0_{a1}0_b0_{a3}0_c\rangle +
|1_{a1}1_b1_{a3}1_c\rangle)
|0_{a2}\rangle /\surd{2},
\end{eqnarray*}
\begin{eqnarray*}
|\phi_4\rangle & = & (|0_{a1}0_b0_{a3}0_c\rangle +
|0_{a1}0_b1_{a3}0_c\rangle + |1_{a1}1_b0_{a3}1_c\rangle -
|1_{a1}1_b1_{a3}1_c\rangle) |0_{a2}\rangle /2\\
& = & [(|0_{a1}0_b0_c\rangle + |1_{a1}1_b1_c\rangle)
|0_{a3}\rangle + (|0_{a1}0_b0_c\rangle -
|1_{a1}1_b1_c\rangle)
|1_{a3}\rangle ] |0_{a2}\rangle /2.
\end{eqnarray*}
\noindent When $M_1 = 0$,

$$|\phi_5\rangle  = (|0_{a1}0_b0_c\rangle +
|1_{a1}1_b1_c\rangle)|0_{a3}\rangle |0_{a2}\rangle
/\surd{2},$$

$$|\phi_6\rangle  = (|0_{a1}0_b0_c\rangle +
|1_{a1}1_b1_c\rangle)
|0_{a3}\rangle |0_{a2}\rangle /\surd{2},$$

$$|\phi_7\rangle  = (|0_{a1}0_b0_c\rangle +
|1_{a1}1_b1_c\rangle)|0_{a3}\rangle |0_{a2}\rangle /\surd{2}.$$

\noindent When $M_1 = 1$,

$$|\phi_5\rangle  = (|0_{a1}0_b0_c\rangle - |1_{a1}1_b1_c\rangle)
|1_{a3}\rangle |0_{a2}\rangle /\surd{2},$$

$$|\phi_6\rangle  = (|0_{a1}0_b0_c\rangle - |1_{a1}1_b1_c\rangle)
|1_{a3}\rangle |0_{a2}\rangle /\surd{2},$$

$$|\phi_7\rangle  = (|0_{a1}0_b0_c\rangle + |1_{a1}1_b1_c\rangle)
|1_{a3}\rangle |0_{a2}\rangle /\surd{2}.$$

\noindent {\it Case 2:} $M_2 =1$
\begin{eqnarray*}
|\phi_3\rangle & = & (|0_{a1}0_b0_{a3}1_{a2}1_c\rangle +
|1_{a1}1_b1_{a3}1_{a2}0_c\rangle)/\surd{2}\\
& = & (|0_{a1}0_b0_{a3}1_c\rangle +
|1_{a1}1_b1_{a3}0_c\rangle)|1_{a2}\rangle /\surd{2},
\end{eqnarray*}
\begin{eqnarray*}
|\phi_4\rangle & = & (|0_{a1}0_b0_{a3}1_c\rangle +
|0_{a1}0_b1_{a3}1_c\rangle + |1_{a1}1_b0_{a3}0_c\rangle -
|1_{a1}1_b1_{a3}0_c\rangle) |1_{a2}\rangle /2\\
& = & [(|0_{a1}0_b1_c\rangle + |1_{a1}1_b0_c\rangle)
|0_{a3}\rangle + (|0_{a1}0_b1_c\rangle - |1_{a1}1_b0_c\rangle)
|1_{a3}\rangle ] |1_{a2}\rangle /2.
\end{eqnarray*}
\noindent When $M_1 = 0$,

$$|\phi_5\rangle  = (|0_{a1}0_b1_c\rangle +
|1_{a1}1_b0_c\rangle)|0_{a3}\rangle |1_{a2}\rangle /\surd{2},$$

$$|\phi_6\rangle  = (|1_{a1}1_b1_c\rangle + |0_{a1}0_b0_c\rangle)
|0_{a3}\rangle |1_{a2}\rangle /\surd{2},$$

$$|\phi_7\rangle  = (|1_{a1}1_b1_c\rangle + |0_{a1}0_b0_c\rangle)
|0_{a3}\rangle |1_{a2}\rangle /\surd{2}.$$

\noindent When $M_1 =1$,

$$|\phi_5\rangle  = (|0_{a1}0_b1_c\rangle -
|1_{a1}1_b0_c\rangle)|1_{a3}\rangle |1_{a2}\rangle /\surd{2},$$

$$|\phi_6\rangle  = (|1_{a1}1_b1_c\rangle - |0_{a1}0_b0_c\rangle)
|1_{a3}\rangle |1_{a2}\rangle /\surd{2},$$
\begin{eqnarray*}
|\phi_7\rangle & = & (- |1_{a1}1_b1_c\rangle
-|0_{a1}0_b0_c\rangle)|1_{a3}\rangle |1_{a2}\rangle /\surd{2}\\
& = & (|1_{a1}1_b1_c\rangle + |0_{a1}0_b0_c\rangle)|1_{a3}\rangle
|1_{a2}\rangle /\surd{2}.
\end{eqnarray*}

The roles of $B$ and $C$ are
symmetrical. Nevertheless, there is a condition on what operations
they should perform when they get a single cbit from $A$. $B$ performs
an $X$ and $C$ performs a $Z$ operation, as required. We set a
cyclic ordering $A \rightarrow B \rightarrow C \rightarrow A$. Let
$A$ be the one sharing EPR pairs with the other two; $A$ is the first
one in the ordering. The second one is $B$, and he must perform an $X$
operation when he gets a single cbit from $A$. The third one is $C$,
and he must perform a $Z$ operation on his qubit when he gets a
single cbit from $A$.  If $B$ is the one sharing EPR pairs with the
other two then $C$ applies an $X$ on his qubit after getting a cbit from
$B$ and, $A$ applies $Z$ on her qubit after getting a cbit from $B$ and
so on.

As we mentioned in the introduction, methods for creating a GHZ
state from Bell pairs have also been discussed by Zukowski et al. 
\cite {zukowski95} and  
Zeilinger et al. \cite {zeilinger97}. First one uses three Bell pairs 
for this purpose, therefore our protocol seems better than theirs in 
the sense that it uses only two Bell pairs. The later, however, uses
only two Bell pairs and only one cbit of communication and seems to
be better than our method at first sight. However, the most interesting 
fact and the motivation for developing our protocol is the dynamic
involvement of both $B$ and $C$ which was lacking in the above
methods. It might me highly desired in many multi-party interactive 
quantum protocols and multi-party cryptograhy (viz. secret sharing)
that both $B$ and $C$ take part actively, say for fairness. 
By fairness we mean that every party has an equal chance for 
participating and effecting the protocol in probabilistic sense. 
It should be interesting to implement this in practical situation.

\section{Preparing a pure $n$-partite maximally entangled state from
EPR pairs shared amongst $n$ agents}
\label{eprsection}

\noindent {\bf Definition 1 ({\it EPR graph}):} 
Suppose there are $n$ agents. We denote them as
$A_1, A_2, ..., A_n$. Construct an undirected graph $G = (V, E)$
as follows:

$$V = \{A_i: i=1, 2, 3,..., n\},$$
$$E = \{\{A_i , A_j \} : A_i ~\mbox{and} ~A_j, ~\mbox{share an EPR pair,}
1\leq i, j\leq n; i \neq j\}.$$

\noindent We call the graph $G = (V, E)$, thus formed, the {\it EPR
graph} of the $n$ agents.

Our definition should not be confused with the {\it entangled graph}
proposed by Plesch and Buzek \cite {plesch02, plesch03}. In entangled 
graph edges represent any kind of entanglement and not neccessarily
maximal entanglement and therefore there is no one to one 
correspondence between graphs and states. EPR graph is unique up to
different EPR pairs. Moreover, we are not concerned with classical
correlations which are also represented by different kind of 
edges in entangled graphs. In an 
EPR graph, two vertices are connected
by an edge if and only if they share an EPR pair.

\noindent {\bf Definition 2 ({\it spanning EPR tree}):}
We call an {\it EPR Graph} $G = (V, E)$ as
{\it spanning EPR tree} when the undirected graph $G = (V, E)$ is a
spanning tree \cite{clr1990}.

We are now ready to develop protocol II to create the $n$-partite
maximally entangled state 
$(|000...0\rangle + |111...1\rangle)/\surd{2}$ along a spanning 
EPR tree. The protocol uses only $O(n)$ cbits of communication and 
local operations. We do not use any qubit communication after
the distribution of EPR pairs to form a spanning EPR tree.


\begin{figure}[htbp]
\centerline{\psfig{figure=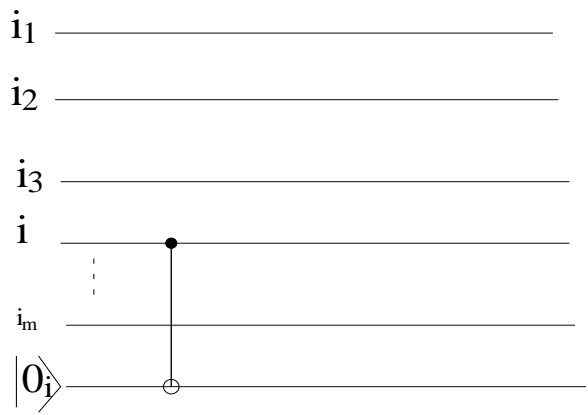,width=10cm,height=12cm}}
\caption{Entangling a new qubit in agent $i$.}
\label{figure5}
\end{figure}

\noindent {\it Protocol II:} Let $G = (V, E)$ be the spanning EPR
tree. Since $G$ is a spanning EPR tree, it must have a vertex say $T
= A_t$ which has degree one (the number of edges incident on a
vertex is called its degree). Note that vertices of $G$ are 
denoted $A_i$, where $1\leq i\leq n$. Let $S = A_s$ be the unique vertex
connected to $T$ by an edge in $G$ and $L$ be the set of all vertices of
$G$ having degree one.  
The vertex $T$ and its only neighbor $S$ are vertices
we start with. We eventually prepare the $n$-partite maximally entangled
state using the three steps summarized below. 
In the first step, one cbit (say 1) is broadcasted by $S$ to
signal other $(n-1)$ parties that the 
protocol for the preparation of the 
$n$-partite entangled state is about to commence.
The second step creates the GHZ state between $S=A_s$, $T=A_t$ 
and another neighbour $R=A_r$ of $S$. The third step 
is the main inductive step where multi-partite entanglement states are
created in a systematic manner over the spanning EPR tree. At the end
of step 3,
when the $n$-partite entangled state is ready, one cbit of 
broadcating from
all the $(k-1)$ elements of $L\setminus \{T\}$ (terminal or degree one 
vertices of $G$) is expected. The $(k-1)th$ such cbit indicates that  
the protocol is over and that maximally entangled state is ready. 
The details are stated below.

\noindent {\it Step 1:} $S$ broadcast one classical bit to signal
the other $(n-1)$ agents that the 
preparation of an $n$-partite entangled state is going to be started
and they must not use their EPR pairs for a qubit teleportation
amongst themselves. In other words, they must save their EPR pairs in
order to use them for the preparation of the $n$-partite entangled
state.

\noindent {\it Step 2:} Clearly $S$ must be connected to a vertex $R$
(say $A_r$) other than $T$ by an edge in $G$, otherwise $G$ will not be
a spanning EPR tree.  A GHZ state among $S$, $T$ and $A_r$ is created.
The GHZ state can be prepared either by using 
usual teleportation circuit, 
the symmetric circuit of protocol I or by the Zeilinger et al. scheme.
Thus using the EPR pairs $(|0_{s,t}0_{t,s}\rangle +
|1_{s,t}1_{t,s}\rangle)/\surd{2}$ and $(|0_{s,r}0_{r,s}\rangle +
|1_{s,r}1_{r,s}\rangle)/\surd{2}$, we prepare the GHZ state
$(|0_{s,t}0_{t,s}0_{r,s}\rangle +
|1_{s,t}1_{t,s}1_{r,s}\rangle)/\surd{2}$. Here the double
subscript $i,j$ denotes that in preparing the given state, EPR pairs
among the agents $A_i$ and $A_j$ have been used.
Here, $A_s=S, A_r=R$ and $A_t=T$.

\noindent {\it Step 3:} Suppose we are currently at vertex $A_i$
and we have already prepared the $m$-partite maximally entangled
state, say

$$(|0_{i1, j1}0_{i2, j2} 0_{i3, j3}...0_{im, jm}\rangle +
|1_{i1, j1}1_{i2, j2}1_{i3, j3}...1_{im,jm}\rangle)/\surd{2},$$

\noindent where $i1 = s, j1 = t, i2 = t, j2 = s, i3 = r, j3 = s$ and
$i = ir$ for some $1 \leq r \leq m$.

\noindent The vertex $A_i$ starts as follows. As soon as he gets
two cbits from one of his neighbors, he completes the operations
required for the success of teleportation and starts processing as
follows. If $A_i \in L$ then $A_i$ broadcast a single cbit.
Otherwise,
(when $A_i \notin L$) let $A_{k1}, A_{k2},..., A_{kp}$
be the vertices connected to $A_i$ by an edge in $G$ such that {\it  k1,
k2,..., kp} are not in the already entangled set with vertex indices 
$\{i1, i2, i3,...,im\}$. $A_i$ takes an extra 
qubit and prepares this qubit in
the state $|0\rangle$ denoted by $|0_i\rangle$. He then prepares
the state

$$(|0_{i1, j1}0_{i2, j2}0_{i3, j3}... 0_{im, jm}0_i\rangle +
|1_{i1, j1}1_{i2, j2}1_{i3, j3}...1_{im, jm}1_i\rangle)/\surd{2}$$

\noindent using the circuit in  Figure \ref{figure5}.
Finally, he teleports his extra qubit to $A_{k1}$ using the EPR
pair $(|0_{i, k1}0_{k1, i}\rangle + |1_{i, k1}1_{k1,
i}\rangle)/\surd{2}$, thus enabling the preparation 
of the $(m+1)$-partite maximally entangled state:

$$(|0_{i1, j1}0_{i2, j2} 0_{i3, j3}...0_{im, jm}0_{k1, i}\rangle +
|1_{i1, j1}1_{i2, j2}1_{i3, j3}...1_{im, jm}1_{k1,
i}\rangle)/\surd{2}.$$

\noindent $A_i$ repeats this until no other vertex, which is
connected to it by an edge in $G$, is left.

\noindent Step 3 is repeated until one cbit each from the
elements of $L$ (except for $T$) is broadcasted, 
indicating that all vertices in $L$ as well
as in $V\setminus L$ have got entangled. 

Note that more than one vertex might be
processing Step 3 at the same time. This however does not matter
since local operations do not change the reduced density matrix of
other qubits. Moreover, while processing the Step 3 together, such vertices will no longer  
be directly connected by an edge in $G$. 

Now we determine the communication complexity of protocol II, 
the number of cbits used in creating 
the $n$-partite maximally entangled state. 
Step 1 involves one cbit broadcast by $S$ to signal the 
initiation of the protocol. To create the GHZ state in Step 2,
atmost $2$ cbits is required. In Step 3, teleportation is 
used to create an $(m+1)$-partite maximally entangled state
from that of $m$-partite. Such $(n-3)$ teleportation steps are
used in this Step entailing $2(n-3)$ cbits of communication.
Finally, $(k-1)$ cbits are broadcast by terminal 
vertices (except $T$). Thus the total cbits used in 
protocol II is 
$1+2+2(n-3)+k-1 = 2n+k-4 \leq 2n+n-1-4 =3n-5=O(n)$.  

Protocol II leads to the following interesting theorem.

\begin{theorem}
If the combinatorial arrangement of distributed EPR pairs 
amongst $n$ agents forms a spanning EPR tree, then the $n$-partite
maximally entangled state $(|000...0\rangle + |111...1\rangle)
/\surd{2}$
can be prepared amongst them with $O(n)$ bits of classical
communication.
\label{qhtheorem1}
\end{theorem}

Theorem \ref{qhtheorem1} thus gives a sufficient condition for preparing  
a maximally entangled $n$-partite state in a distributed network
of EPR pairs. In order to prove this sufficiency, 
we have also developed two more protocols 
which require $O(n)$ cbits of communication. 
The first two steps of these protocols are essentially the 
same as that of Protocol II.
The first protocol involves all the 
agents already entangled in each iteration in Step 3, where, a circuit
very similar to the symmetric teleportation circuit 
(Figure \ref{figure3}) of Protocol I is used. The classical communication 
cost is $(2n-4)$ bits. The second protocol 
uses a generalization of the method of  
Zeilinger et al. in each iteration of Step 3 and requires   
$(2n-3)$ cbits of communication. In this 
paper we have presented only Protocol II 
instead of these two protocols because 
of simplicity and the direct use of teleportation.\\

The question of interest now is that of determining 
the minimal structure 
or combinatorics of the distribution of EPR pairs neccessary for
creating the $n$-partite maximally entangled state. 
In other words, we wish to characterize 
neccessary properties to be satisfied by the EPR graph
for this purpose. We argue below that the EPR graph, indeed,
must contain a spanning EPR tree, and must therefore be connected. 
We assume for the sake of contradiction that the EPR graph $G$ is
not connected. 
Then, it must have at least two components,
say $C_1$ and $C_2$. No member of $C_1$ is connected to any member of 
$C_2$ by an edge in $G$. This means that no member of $C_1$ is
sharing an EPR pair with any member of $C_2$. 
Suppose a protocol $P$ can create a pure $n$-partite maximally 
entangled state starting from the disconnected EPR graph $G$
of $n$ agents. 
If we are able
to create an $n$-partite maximally entangled state using
protocol $P$ with this 
structure using only classical communication and local operations,
it is easy to see that we will also be able to create 
an EPR pair between two parties that were not earlier sharing 
any EPR pair, using just local operations and classical communication.
This can be done as follows. 
Let $A$ be the first party that 
posseses all the qubits of his group (say $C_1$)
and $B$ be the second party that possesses all the qubits
of his group (say $C_2$). Now the protocol $P$ is run on this
structure to create the $n$-partite maximally entangled state. 
Then, $A$ ($B$) disentangles all of his 
qubits except one by reversing the circuit in Figure \ref{figure5};
this leaves $A$ and $B$ sharing an EPR pair.
This means that two parties which were never sharing an EPR pair
are able to share it just by local operations 
and classical communication (LOCC).
This is forbidden by  
fundamental laws in quantum information theory (LOCC cannot increase the
expected entanglement \cite{vedral02}), hence $G$ must be connected. 
Note that no qubit communication is permitted 
after the formation of EPR graph $G$.  
We present this neccessary condition in the following theorem.

\begin{theorem}
A necessary condition that the $n$-partite maximally entangled state 
$(|000...0\rangle + |111...1\rangle)/\surd{2}$ be prepared
in a distributed 
network permitting only EPR pairs for pairwise entanglement between agents 
is that the EPR graph of the $n$ agents must be connected.
\label{qhtheorem2}
\end{theorem}

It can be noted that after the preparation of the state 
$(|000...0\rangle + |111...1\rangle)/\surd{2}$, any other
pure $n$-partite maximally entangled state can also be
prepared by just using local operations.
We also know that any connected undirected graph contains a 
spanning tree \cite{clr1990}. Thus a connected EPR graph 
will contain a spanning EPR tree.
With this observations,
we combine the above two theorems in the following theorem.

\begin{theorem}
Amongst $n$ agents in a communication network permitting only 
pairwise entanglement in the form of EPR pairs, a pure $n$-partite
maximally entangled state can be prepared if and only if the 
EPR graph of the $n$ agents is connected. 
\label{qhtheorem3}
\end{theorem}

\section{Entangling a set of agents from entangled states of subsets: Combinatorics
of general entanglement structure }
\label{hypersection}

In the previous section we have presented the necessary and sufficient
condition for preparing a pure multi-partite maximally entangled state in a 
distributed network of EPR pairs (see Theorem \ref{qhtheorem3}). 
However, agents 
may not be connected by EPR pairs in a general network. We assume that 
subsets of agents may be sharing pure maximally entangled states. So, some 
triples of agents may be GHZ entangled, some pairs of agents may share EPR 
pairs and some subsets of agents may share even higher dimensional entangled
states. \\

Now we develop the combinatorics of multi-partite entanglement within subsets 
of agents required to 
prepare multi-partite entanglement between all the agents. When we were 
dealing only with EPR pairs in the case of EPR graphs or spanning EPR trees,
we used the simple graph representation. Now subsets of the set of all 
agents may be in multi-partite entangled states and therefore we use a natural 
representation for such entanglement structures with hypergraphs as follows.\\

Let $S$ be the set of $n$ agents in a 
communication network. Let $E\subset S$, 
$|E|=k$. Suppose $E$ is such that the $k$ agents in $E$ are 
in a $k$-partite
pure maximally entangled state.
Let $E_1$, $E_2$, ..., $E_m$ be such subsets of $S$, each having a 
pure maximally entangled shared state amongst its agents.
Note that the sizes of these subsets may be different.
Consider the
hypergraph $H=(S,F)$ \cite{hcvol1} such 
that $F=\{E_1,E_2,...,E_m\}$. We call
such a
hypergraph $H$, an {\it entangled hypergraph}
of the $n$ agents. In standard hypergraph notation the elements of $F$ are 
called {\it hyperedges} of $H$.
Now we present the necessary and sufficient condition for
preparing a $n$-partite pure maximally entangled state in such 
networks, given entanglements as per the entangled hypergraph.
We need the definition of a {\it hyperpath} in a hypergraph: a sequence of
$j$ hyperedges $E_1$, $E_2$, ..., $E_j$ in a hypergraph
is called a {\it hyperpath} from a vertex $a$ to a vertex $b$ 
if (i) $E_i$ and $E_{i+1}$ have a common vertex (agent)
for all $1\leq i\leq j-1$ (ii) $a$ and $b$ are agents in $S$ 
(iii) $a\in E_1$ and (iv) $b\in E_j$. If there is a hyperpath between every
pair of vertices of $S$ in a hypergraph $H$ then we say that $H$ is 
connected.  

\begin{theorem}

Given $n$ agents in a communication network and an entangled hypergraph, a
pure $n$-partite maximally entangled state can be
prepared amongst the $n$ agents if and only if the entangled hypergraph 
is connected.

\label{entanhyperiff}

\end{theorem}

The proof of this theorem is based on the following Protocol III.

\noindent {\it The Protocol III:}
We assume without loss of generality that $n>|E_1|\geq |E_2|...\geq |E_m|$.
We maintain the set $F$ and $R=S\setminus F$ where $F$ contains 
the agents already entangled in the $|F|$-partite pure maximally 
entangled state. Initially, $F=E_1$ and $R=\{E_2, E_3,..., E_m\}$.
We repeat the following steps until $F=S$. 
Choose $E_i\in R$ with minimum $i$ such that $F$ and $E_i$ have at least one
common agent and $E_i$ is not in $F$; 
let the smallest index common element between $E_i$ and $F$ 
be the agent $A_j$. (Since the entangled hypergraph 
is connected, there is always such a hyperedge $E_i$.) We 
can now use the method of Zeilinger et al. to create an 
$(N+M-1)$-partite maximum entangled state from two groups, one 
containing $N=|F|$ agents
and the other containing $M=|E_i|$ agents. 
The measurement is processed by $A_j$. 
So, an $(|F|+|E_i|-1)$-partite
entanglement state is prepared from amongst the members of $F$ and $E_i$.
If $F$ and $E_i$ share only one common agent then we are done. Otherwise, each 
member common to $F$ and $E_i$ other than $A_j$ will have two qubits each
from the $|F|+|E_i|-1$ entangled qubits.
These qubits must be disentangled using a circuit same as the reverse of
circuit in Figure \ref{figure5}. Now the members of $F$ and $E_i$ 
remain entangled in $(|F|+|E_i|-|F \bigcap E_i|)$-partite state, 
each holding 
exactly one qubit. Finally, we set $F=F \bigcup E_i$ and $R=R\setminus E_i$.

The proof of necessity is similar to that of
the proof of necessity in Theorem \ref{qhtheorem2}. 
For the sake of contradiction assume that the entangled hypergraph $H$
is not connected. Then there is no hyperpath between two agents (say 
$a$ and $b$), implying the existence 
of at least two components $C_1$ and $C_2$
in $H$, with no member of $C_1$ sharing 
a hyperedge of entanglement with any 
member of $C_2$. 
Suppose a protocol $P$ can create a pure $n$-partite maximally 
entangled state starting from the disconnected entangled hypergraph 
$H$ of $n$ agents. 
If we are able
to create an $n$-partite maximally entangled state using
protocol $P$ with this 
structure using only classical communication and local operations,
it is easy to see that we will also be able to create 
an EPR pair between two parties that were not earlier sharing 
any EPR pair, using just local operations and classical communication.
This is forbidden by 
fundamental laws in quantum information theory 
(LOCC cannot increase the
expected entanglement \cite{vedral02}). Hence $H$ must be connected. 
This completes the proof of Theorem \ref{entanhyperiff}.

\section{Concluding remarks}
\label{conclusion}

We compare our method (Protocol II) of
generating multipartite maximally entangled states
with that of Bose et al. \cite{bose98}.
The scheme of Bose et al. works as follows.
Each agent needs to share a Bell pair with a 
central exchange 
in the communication network of $n$ agents.
The central exchange then projects the $n$-qubits 
with him, on to the $n$-partite maximally entangled basis.
This leaves the $n$ agents in a $n$-partite maximally 
entangled state. Thus, the two basic requirements 
of their scheme are a central exchange and a projective 
measurement on a multi-partite maximally entangled basis.
The central exchange essentially represents a star topology 
in a communication network and allows certain degree of 
freedom to entangle particles belonging to any set of users
only if the necessity arises. 
However, a real time communication network may not
always be a star network, in which case, we may need to
have several such central exchanges. 
Of course, one will also be interested in setting up
such a network with minimum resources, especially in 
the case of a 
long distance communication network. 
Issues involved in the design of such 
central exchanges such as 
minimizing required resources, are of vital interest
while dealing with real communication 
networks. Such networks may be called 
{\it Quantum Local Area Network} (Q-LAN or Non-LAN, a Non-Local LAN)
or {\it Quantum Wide Area Network} (Q-WAN or Non-WAN).
Our scheme presented in Section \ref{eprsection} adresses these  
issues.
We have shown in Theorem \ref{qhtheorem3}
that the {\it spanning EPR tree} is 
the minimal combinatorial requirement for this purpose.
The star topology is a special case of the spanning EPR tree where
the central exchange is one of the agents.
It is therefore clear that 
the star network requirements of the scheme of
Bose et al. provides a sufficient condition where as the 
requirement in our spanning EPR tree
scheme is the most general and minimal possible structure.\\

Our scheme also helps in minimizing resources.
Our spanning tree topology has been used by 
Singh and Srikanth \cite{sinsrik031} for this purpose.
They assign weights to the edges of the EPR graph
based on the resources (such as quantum repeaters, etc.)
needed to build that particular edge. 
Then, a {\it minimum spanning EPR tree} represents 
the optimized requirement.
They also use this topology for multi-party quantum cryptography
to minimize the size of the sector that can be potentially
controlled by an evesdropper. 
Thus our topology seems to be a potential candidate for
building a long distance quantum communication network
(such as in a Non-LAN or Non-WAN).\\

The second basic ingredient of the scheme of Bose et al. is 
the projection on a multipartite maximally entangled basis.
As they point out, the circuit for such a measurement 
is an inverse of the circuit that generates a maximally 
entangled state from a disentangled input in the 
computational basis. 
In a communication network involving a large number
of agents, this entails a lot of work to be done
on part of the central exchange while the agents are
idle.
In our scheme, work is distributed 
amongst the agents.
Moreover, the $n$-qubit joint measurement on the entangled basis 
in the scheme of Bose et al. seems to be well high impossible from 
a practical standpoint given the current technology, 
whereas all the practical requirements of our scheme (Protocol II)  
can be met using current technology (using telecom cables to distribute
entanglement etc).\\  

The projection used by the central exchange in the 
scheme of Bose et al. may lead to any of the 
$2^n$ possible $n$-partite maximally entangled states.
For practical purposes, one might be more interested
in a particular state. 
To get the desired state, the measurement result 
must be broadcast by the central exchange.
The $2^n$ possible states can be represented by 
a $n$ bit number and thus the communication 
complexity involved in their scheme is $n$ 
cbits, essentially the same as that of ours asymptotically.
Therefore, our scheme is comparable to their scheme also
in terms of communication complexity.
It can also be noted at this point that,
in our topology, even the method of
Zeilinger et al. for creating $(m+1)$-partite
maximally entangled state from a $m$-partite
maximally entangled state becomes applicable.
The use even reduces the communication complexity
by some cbits but still requires $2n-3$ cbits which is $O(n)$.
As it can be observed, all these schemes require 
$O(n)$ cbits of communication. 
Whether there is an $\Omega (n)$ lower bound 
on the cbit communication complexity for preparing 
$n$-partite a pure maximally entangled state given a 
spanning EPR tree
remains open for further research.\\

The results in Theorem \ref{qhtheorem3} and 
Theorem \ref{entanhyperiff} are similar to the
classical theorem by Helly \cite{v1964} in convex geometry.  Helly's
theorem states that a collection of closed convex sets in the
plane must have a non-empty intersection if each triplet of the
convex sets from the collection has a non-empty intersection. In
one dimension, Helly's theorem ensures a non-empty intersection of
a collection of intervals if each pair of intervals has a
non-empty intersection. In our case (Theorems \ref{qhtheorem3} and 
\ref{entanhyperiff}), there is
similar combinatorial nature; if $n$ agents are such that each pair
has a shared EPR pair, then (with linear classical communication
cost) a pure $n$-partite state with maximum entanglement can be created
entangling all the $n$ agents. As stated in Theorem \ref{qhtheorem1},
the case is
stronger because just $(n-1)$ EPR pairs suffice. Due to this
similarity in combinatorial nature, we call our results in Theorem
\ref{qhtheorem3} and Theorem \ref{entanhyperiff} 
quantum Helly-type theorems.\\

{Acknowledgments:}
We thank R. Srikanth, K. Mitra and S. P. Khastigir for a thorough reading of the manuscript and
for their comments.

\end{document}